\batchmode
\makeatletter
\def\input@path{{/home/flatmax/flatmax/personal.work/research/zeroPhaseResearch/impulseModel/}}
\makeatother
\documentclass[australian]{article}
\usepackage[T1]{fontenc}
\usepackage[latin9]{inputenc}
\usepackage{geometry}
\usepackage{amsmath}
\usepackage{graphicx}
\usepackage{babel}
\begin{document}
\title{Allpass impulse response modelling}
\author{Matthew R Flax <flatmax@flatmax.org>}

\maketitle
This document defines a method for FIR system modelling which is very
trivial as it only depends on phase introduction and removal (allpass
filters). As magnitude is not altered, the processing is numerically
stable. It is limited to phase alteration which maintains the time
domain magnitude to force a system within its linear limits.

\section{Introduction}

A trivial method for capturing the impulse response of a system is
to force it with impulses. The captured response is the impulse model
(impulse response) of the system. In practice forcing systems with
impulses either requires high energy signals or a very long time to
stack captures to reduce noise. Problems with using high energy signals
is that the systems become nonlinear and the impulse models do not
represent the linear components of system accurately.

By spreading the signal energy out in time, other common approaches
are to use uncorrelated noise signals or chirping sinusoids. The captured
output of the system from these forcing signals is then processed
in magnitude and phase to generate the impulse model of the system.
For example the Levinson-Durban algorithm is an auto-regressive process
which uses signal correlations and cross correlations to generate
the impulse model \cite{Proakis:Book}. For a long impulse model,
the Levinson-Durban algorithm becomes computationally complex and
takes a long time.

Another direct method for impulse model generation is to simply divide
the captured signal (in the Fourier domain) from the reference signal
which is a simple form of deconvolution. This simple method of deconvolution
is numerically unstable for signals with no power at particular frequencies.

This document introduces a simple method for finite impulse response
(FIR) system identification. It is a very simple approach and is numerically
stable. The approach uses a band limited impulse (Section \ref{sec:Zero-phase-impulse})
and applies an allpass filter to generate a linear chirp (Section
\ref{sec:Allpass-chirping-filter}). A system is forced with this
linear chirp (Section \ref{sec:Forcing-the-system}) and the impulse
model of the system is produced by removing the phase of the allpass
filter (\ref{sec:Generating-the-impulse}). As magnitude is not used
in the processing, it is numerically stable.

The accuracy of the technique presented in this document is limited
by the noise level in the captured signal, as noise is not suppressed
by anything other then stacking.

\section{Zero phase impulse\label{sec:Zero-phase-impulse}}

\begin{figure}
\begin{centering}
\includegraphics[width=0.45\linewidth]{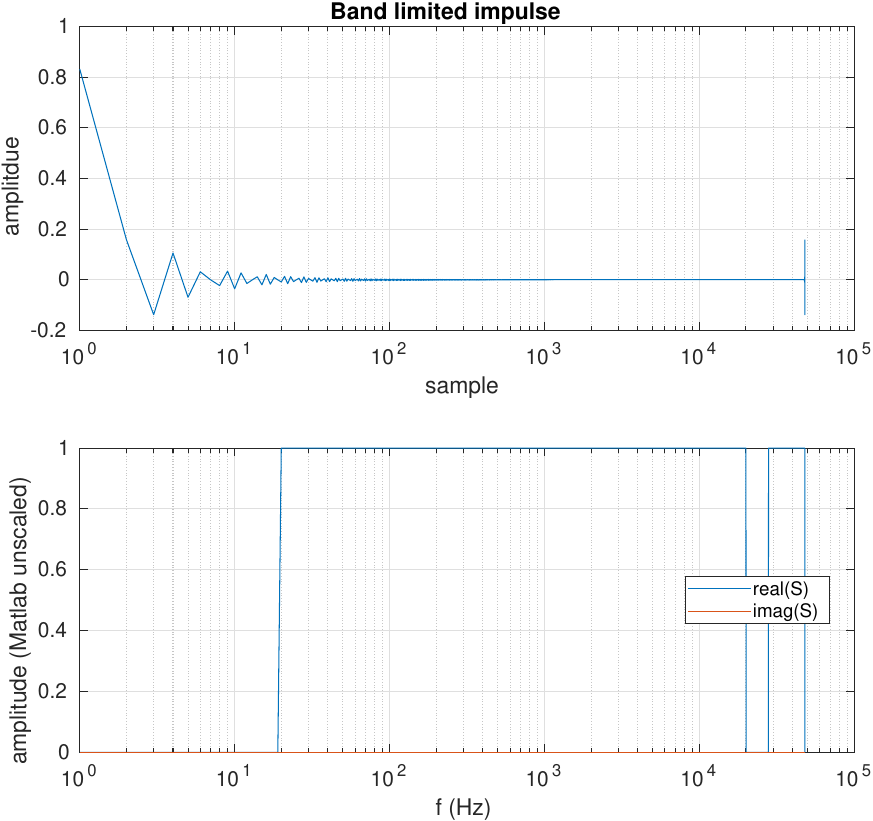}
\par\end{centering}
\caption{Band limited impulse\label{fig:Band-limited-impulse}}
\end{figure}

A zero phase band limited impulse ($\Delta$, with minimum frequency
($f_{i}$) and maximum frequency ($f_{a}$)) is defined for frequencies
between 0 and the half the sample rate ($f_{s}$)
\begin{align}
\Delta_{f_{i},f_{a}} & =\begin{cases}
0 & 0\le f<f_{i}\\
2N^{-1} & f_{i}\le f\le f_{a}\\
0 & f_{a}<f\le\frac{f_{s}}{2}
\end{cases}\label{eq:bl-imp}
\end{align}
where N is the number of samples in the signal and $\Delta$ is the
discrete Fourier transform of $\delta$\footnote{In this document lower case symbols represent signals in the time
domain and upper case symbols represent signals in the discrete Fourier
domain}. This band limited impulse is shown in Figure \ref{fig:Band-limited-impulse}.

\section{Allpass chirping filter\label{sec:Allpass-chirping-filter}}

\begin{figure}
\begin{centering}
\includegraphics[width=0.45\linewidth]{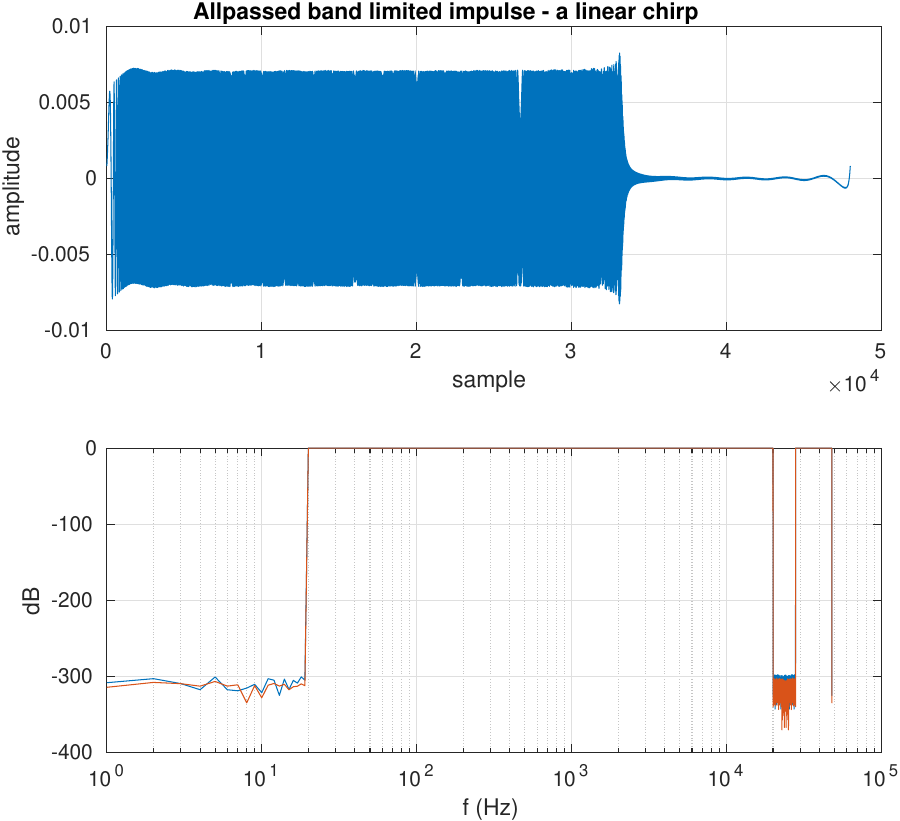}
\par\end{centering}
\caption{A linear chirp formed by allpass filtering a band limited impulse.\label{fig:bl-chirp}}
\end{figure}

The band limited zero phase impulse $\Delta$ is filtered to a linear
chirp using the relationship between instantaneous frequency and phase
($\phi=2\pi\int fdn$)
\begin{align}
f_{c} & =\frac{1}{2T}\left(\left(f_{a}-f_{i}\right)t+f_{i}\right)\nonumber \\
\phi_{c} & =2\pi\int\frac{1}{2T}\left(\left(f_{a}-f_{i}\right)t+f_{i}\right)dt\nonumber \\
 & =\frac{\pi}{T}\left(\left(f_{a}-f_{i}\right)t^{2}+f_{i}t\right)+c\label{eq:all-pass-phase}
\end{align}
where T is the total duration of the signal and $c$ is a constant.
The definition of the reference filter (r) is the convolution of the
band limited impulse (Equation \ref{eq:bl-imp}) with the all pass
filter (phase specified in Equation \ref{eq:all-pass-phase})
\[
R=\Delta_{f_{i},f_{a}}e^{j\phi_{c}}
\]
shown in Figure \ref{fig:bl-chirp}

\section{Forcing the system\label{sec:Forcing-the-system}}

\begin{figure}
\begin{centering}
\includegraphics[width=0.45\linewidth]{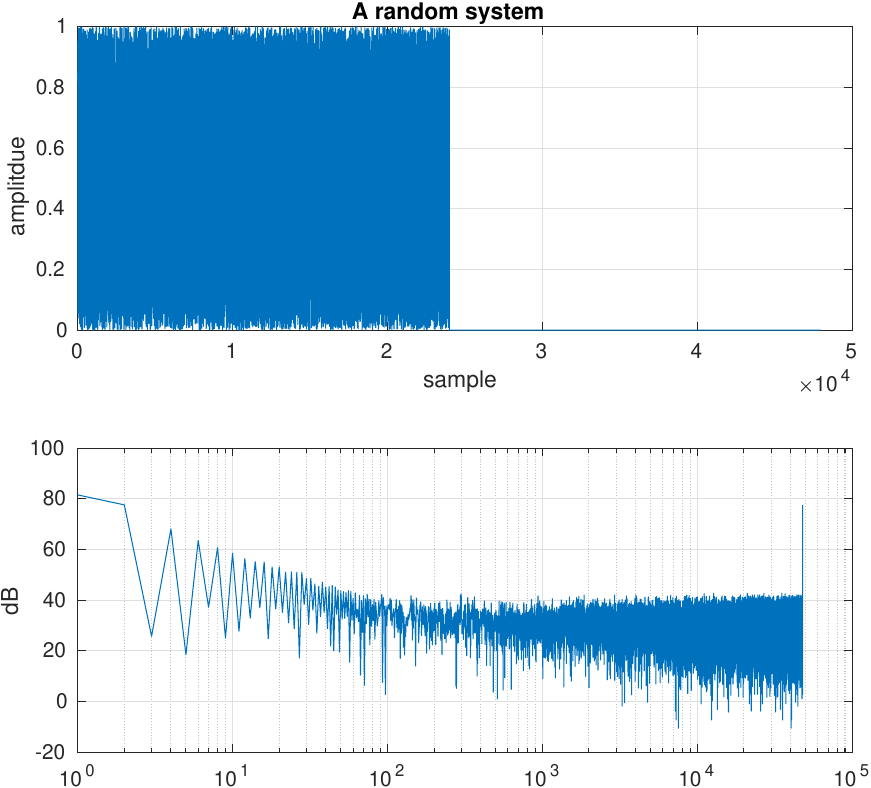}
\par\end{centering}
\caption{A random wide band system used in this theoretical experiment.\label{fig:system}}
\end{figure}

A system is forced by the reference signal (r). In this theoretical
treatment, a random wide band system is used (h), shown in Figure
\ref{fig:system}. The forced system response (y) is defined as
\[
Y=HR
\]

\section{Generating the impulse model\label{sec:Generating-the-impulse}}

\begin{figure}
\begin{centering}
\includegraphics[width=0.45\linewidth]{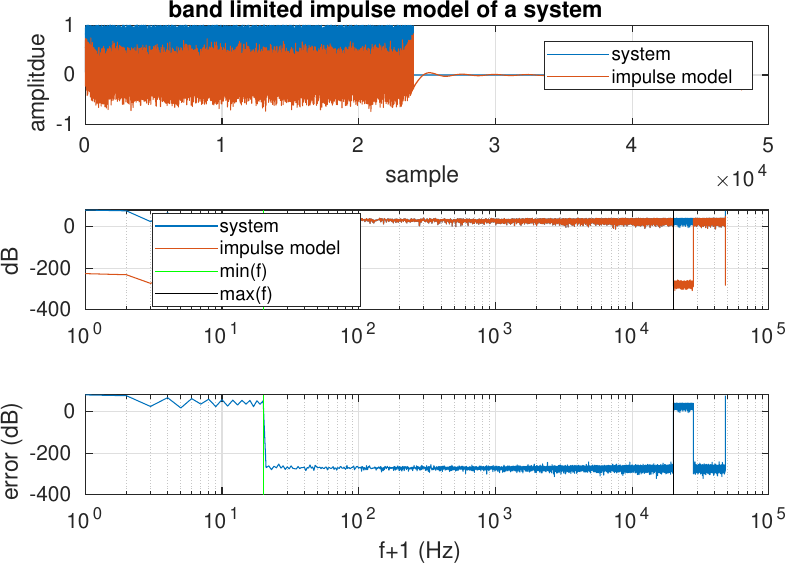}
\par\end{centering}
\caption{The narrow band impulse model of the wideband system. The blue line
is the system from Figure \ref{fig:system} and the red line is the
impulse model of the system found using Equation \ref{eq:impulse-model}.\label{fig:impulse-model}}
\end{figure}

To find the impulse model ($\hat{h}$), the forced system (y) is filtered
by the inverse allpass filter
\begin{align}
\hat{H} & =Ye^{-j\phi_{c}}\label{eq:impulse-model}\\
 & =HRe^{-j\phi_{c}}\nonumber \\
 & =H\Delta_{f_{i},f_{a}}e^{j\phi_{c}}e^{-j\phi_{c}}\nonumber \\
 & =H\Delta_{f_{i},f_{a}}\nonumber 
\end{align}
which is the system convolved by a bandlimited impulse (in other words,
the band limited system itself), shown in Figure \ref{fig:impulse-model}.
The error between the system and the impulse model of the system is
shown by the bottom image in Figure \ref{fig:impulse-model}. There
is very little error in the band of the impulse, outside the band
of the impulse, the system is not defined. This highlights the power
of using only phase and not magnitude in the computation, the resulting
impulse model is numerically stable.

\section{Frequency doubling}

{[}take 1{]}Exponential sweeping of the frequency from DC is defined
as
\begin{align*}
\omega_{c} & =\omega_{i}2^{\left(\frac{t}{T}log_{2}\frac{\left(\omega_{a}\right)}{\left(\omega_{i}\right)}\right)}\\
 & =\omega_{i}\left(\frac{\omega_{a}}{\omega_{i}}\right)^{\frac{t}{T}}\\
\phi_{c} & =\int\omega_{c}dt\\
 & =\frac{\omega_{i}T\left(\frac{\omega_{a}}{\omega_{i}}\right)^{\frac{t}{T}}}{log\left(\frac{\omega_{a}}{\omega_{i}}\right)}+c
\end{align*}
where the constant c can be set to keep the phase zero at time zero
\[
c=-\frac{\omega_{i}T}{log\left(\frac{\omega_{a}}{\omega_{i}}\right)}
\]

In order to keep the amplitude constant for all frequencies a new
amplitude scaling term ($a_{c}$) is introduced
\begin{align*}
a_{c} & =\left(\frac{d\omega_{c}}{dt}\right)^{-1}\\
 & =\frac{\omega_{i}log\left(\frac{\omega_{a}}{\omega_{i}}\right)\left(\frac{\omega_{a}}{\omega_{i}}\right)^{\frac{t}{T}}}{T}
\end{align*}

{[}take 2{]} Exponential sweeping of the frequency from DC is defined
as
\begin{align*}
\omega_{c} & =2^{\left(\frac{t}{T}log_{2}(\omega_{a}+1)\right)}-1\\
 & =(\omega_{a}+1)^{\frac{t}{T}}-1\\
\phi_{c} & =\int\omega_{c}dt\\
 & =\frac{T\left(\omega_{a}+1\right)^{\frac{t}{T}}}{log\left(\omega_{a}+1\right)}-t+c
\end{align*}
where the constant c can be set to keep the phase zero at time zero
\[
c=-\frac{T}{log\left(\omega_{a}+1\right)}
\]

In order to keep the amplitude constant for all frequencies a new
amplitude scaling term ($a_{c}$) is introduced
\begin{align*}
a_{c} & =\left(\frac{d\omega_{c}}{dt}\right)^{-1}\\
 & =\frac{T}{log\left(\omega_{a}+1\right)}\left(\omega_{a}+1\right)^{-\frac{t}{T}}
\end{align*}

{[}This is old{]} Exponential sweeping of the frequency is defined
as
\begin{align*}
f_{c} & =f_{i}2^{\left(\frac{t}{T}log_{2}\left(\frac{f_{a}}{f_{i}}\right)\right)}\\
\phi_{c} & =2\pi\int f_{c}dt\\
 & =\frac{2\pi Tf_{i}}{log\left(\frac{f_{a}}{f_{i}}\right)}\left(\frac{f_{a}}{f_{i}}\right)^{\frac{t}{T}}+c
\end{align*}
In order to keep the amplitude constant for all frequencies a new
amplitude scaling term ($a_{c}$) is introduced
\begin{align*}
a_{c} & =a_{i}\frac{df_{c}}{dt}\\
 & =a_{i}\frac{f_{i}log\left(\frac{f_{a}}{f_{i}}\right)}{T}\left(\frac{f_{a}}{f_{i}}\right)^{\frac{t}{T}}
\end{align*}

\section{Conclusion}

This document defines a method for FIR system modelling which is very
trivial as it only depends on phase introduction and removal (allpass
filters). As magnitude is not altered, the processing is numerically
stable. It is limited to phase alteration which maintains the time
domain magnitude to force a system within its linear limits.

The approach starts with a band limited double sided impulse (zero
phase - see Section \ref{sec:Zero-phase-impulse}). The impulse is
filtered (using an allpass filter) to produce a linear chirp, in Section
\ref{sec:Allpass-chirping-filter}, see Figure \ref{fig:bl-chirp}.
This chirp is used to force a system (in our example a random system
- shown in Figure \ref{fig:system}). Finally to find the impulse
model of the system, an inverted phase allpass filter is used to produce
a band limited model of the system (see Figure \ref{fig:impulse-model}).

The resulting system model captures both the phase and magnitude of
the system, but only within the specified frequency limits of the
band limited impulse.

\bibliographystyle{plain}
\bibliography{4_home_flatmax_flatmax_personal_work_research_XDerivations_biblography}

\end{document}